\documentclass[journal]{IEEEtran}

\usepackage{graphicx} 

\begin{document}

\title{DFT Modeling of Bulk-Modulated Carbon Nanotube Field-Effect Transistors}

\author{Luca~Latessa,~Alessandro~Pecchia,~and~Aldo~Di~ Carlo%
\thanks{L. Latessa, A. Pecchia and A. Di Carlo are with the University of Rome ``Tor Vergata'', Department of Electronic Engineering, Rome 00133, Italy (e-mail: latessa@ing.uniroma2.it)}}


\maketitle

\begin{abstract}
We report density-functional theory (DFT), atomistic simulations of the non-equilibrium transport properties of carbon nanotube (CNT) field-effect transistors (FETs). 
Results have been obtained within a self-consistent approach based on the non-equilibrium Green's functions (NEGF) scheme. 
Our attention has been focused on a new kind of devices, the so called bulk-modulated CNTFETs. 
Recent experimental realizations \cite{Chen,Lin_condMat} have shown that such devices can exhibit excellent performances, even better than state-of-the-art Schottky barrier (SB)-modulated transistors. 
Our calculations have been intended to explore, at an atomistic level, the physical mechanisms governing the transport in these new devices.
We emphasize the role that one-dimensional screening has on gate- and drain-induced current modulation mechanisms, pointing out, at the same time, the importance of a correct evaluation of the nanotube quantum capacitance.
The operative regimes and the performance limits of the device are analysed, pointing out, at the same time, the role played by the quasi-one-dimensional, short channel effects.

\end{abstract}


%
\IEEEpeerreviewmaketitle

\section{Introduction}

\PARstart{C}{arbon} nanotubes have been recently acknowledged as promising candidates, among several low-dimensional physical systems, to realize nanoscale electronic devices \cite{McEuen,Appenzeller04}. 
In particular, single-wall CNTs, consisting of a single graphene sheet wrapped up to form tubes with very small diameters (up to $\sim$ $1$ $nm$), are ideal candidates to study general transport properties of quasi-one dimensional devices.

CNTs posses exceptional electrical and mechanical properties, such as an exceptional strength and stability, which generate in turn the capability of carrying very high current densities \cite{Yao}; very large values of the mean free path have been found, which guarantee carriers to propagate ballistically over very long distances even at room temperature ($\sim$ $1$ $\mu m$ in the low field regime) \cite{Park}.
A CNT can be either metallic or semiconducting, depending on its chirality.
Metallic CNTs can be employed as good interconnections among electrical devices at the nanoscale.
Semiconducting CNTs have been instead used to demonstrate field-effect transistors having quasi-one dimensional channels \cite{Tans}.

The simplest approach to realize a nanotube transistor consists of contacting a semiconducting nanotube with two metallic contacts acting as source and drain. 
A gate modulates the injection of carriers in the nanotube channel by modulating the barrier height at the metal-semiconductor interface. 
A Schottky barrier (SB) nanotube transistor results in this case \cite{Heinze}. 
Significant progresses have been made in the fabrication and the understanding of these kind of devices.
A SB-CNT transistor is essentially governed by a contact-dominated switching. This is the fundamental difference with respect to conventional MOSFET devices, governed by conductance modulation of the channel. 
Indeed, SB-modulation strongly affects the performances of the nanotube device. 
The contact-dominated control mechanism leads to a large sub-threshold swing $S=dV_{G}/d(logI_{DS})$ (where $V_{G}$ and $I_{DS}$ are respectively the gate voltage and the source-drain current), substantially limiting the gate insulator thinning, since a strong ambipolar behavior appears when the device is vertically scaled \cite{Heinze,Radosav,Guo_SB}.
These drawbacks can prevent nanotube transistors from holding particular promises for high-performance nano-electronics.

Recently, novel device configurations for bulk-switched CNTFETs have been proposed and demonstrated \cite{Chen,Lin_condMat}.
To obtain a channel-dominated behavior, a local gate contact is realized in the middle point of the channel, strongly reducing electrostatic coupling with the source and drain contacts \cite{Lin}.
The height of the contacts barrier simply fixes the charge injected in the channel, regardless the value set for the gate bias. The carrier injection can be controlled either by electrostatic or chemical doping of the ungated portion of the nanotube \cite{Lin_condMat}.

This device scheme allows to greatly improve the nanotube transistor performances. A transition from ambipolar to unipolar behavior has been demonstrated. Moreover, measured devices have shown a very small sub-threshold swing, very close to the theoretical limit found for standard MOSFET, and an excellent $I_{on}/I_{off}$ ratio.    
         
Nevertheless, important issues need still to be addressed for the bulk-switched CNTFET.
The conductance modulation mechanism involved in the device behavior is obtained by locally modulating the barrier height for carrier transmission through the nanotube.
Consequently, local screening properties of the quasi-one dimensional electron gas associated to the nanotube can assume a predominant role in the switching mechanism.

In this work we investigate, at an atomistic level, the physical mechanisms ruling current flow in a coaxially gated CNTFET. The role of the nanotube screening properties on transport has also been pointed out.
To correctly describe screening, calculation have been performed within a DFT-based, self-consistent approach, on which the NEGF technique has been implemented.


\section{Formalism and methodology} 


All results presented in this work have been obtained by using the quantum transport simulator gDFTB \cite{Review,Book,DiCarlo}.
In the gDFTB formalism, the non-equilibrium Green's functions scheme is implemented on a self-consistent, DFT-based tight-binding framework.

The density-functional Hamiltonian matrix elements are evaluated starting from the Slater-Koster two-center approximation \cite{Slater}. 
These elements, represented on a suitable non-orthogonal atomic-like basis set, are obtained by first principle computations as a function of atomic distance and are stored. Starting from this parametrization, the matrix elements for the actual system to be simulated are simply obtained by interpolation.
This approach combines an \emph{ab initio} accuracy with highly efficient atomistic computations. 

The tight-binding Hamiltonian is evaluated self-consistently with the local charge fluctuation induced on the system by charge transfer among atoms \cite{Elstner}. 
The exchange and correlation (XC) contributions to the DFT energy functional are considered in a local density approximation (LDA), and are included in the self-consistent Hamiltonian as on-site Hubbard energies \cite{Elstner}. 
Details of this self-consistent, DFT-based approach and its successful applications to a large variety of organic and inorganic systems have been extensively given elsewhere \cite{Fraunheim}.

In order to properly treat charge transfer taking place when current flows through the system, we calculate the non-equilibrium carriers distribution by employing the NEGF technique.
The occupation of the non-equilibrium scattering states of the system is completely resolved by knowing the density matrix $\rho_{\mu\nu}$.
In the NEGF formalism, this fundamental quantity can be obtained starting just from the non-equilibrium Green's function of the system, $G^{<}_{\mu\nu}$, as \cite{Datta}
\begin{equation} \label{eq:densmx}
\rho_{\mu\nu}=\frac{1}{2\pi i}\int^{+\infty}_{-\infty}G^{<}_{\mu\nu}(E)dE.
\end{equation}
We evaluate this integral by using standard methods proposed in the literature \cite{Brandbyge}.
Technical details more specific to our approach can be found in Ref. \cite{Book}.
Here we only limit ourselves to give some details on $G^{<}_{\mu\nu}$, in order to give an intuitive insight in the physical meaning of the formalism.
In its most general formulation, the steady-state non-equilibrium Green's function of the system is given by the Keldish-Kadanoff-Baym (KKB) equation \cite{Kadanoff,Keldysh,Danielewicz},
\begin{equation} \label{eq:greenf}
G^{<}(E)=G^{R}(E)\Sigma^{<}(E)G^{A}(E),
\end{equation}
where $G^{R(A)}=\left[ES-H-\Sigma_{1}^{R(A)}-\Sigma_{2}^{R(A)}\right]^{-1}$ is the standard, equilibrium, retarded (advanced) Green's function of the system, and $\Sigma^{<}$ is the non-equilibrium self-energy.
The quantities which appear in the expression of $G^{R(A)}$ are the self-consistent, DFT Hamiltonian matrix $H$, the overlap matrix $S$, resulting from the choice of a non-orthogonal tight-binding basis set, and the self-energies of the two contacts, $\Sigma_{1.2}^{R(A)}$, describing the coupling of the device with the two semi-infinite reservoirs \cite{Datta}.   
Scattering sources, such as electron-phonon or electron-electron interaction, can be included in the general expression of $\Sigma^{<}$ to describe incoherent transport.
The method we use in this work is restricted to the description of coherent transport mechanisms. 
In this case, no scattering mechanisms are included, and $\Sigma^{<}$ is determined only by the coupling of the device with the two semi-infinite contacts, resulting $\Sigma^{<}(E)=-2iIm\left[f_{1}\Sigma_{1}^{R}(E)+f_{2}\Sigma_{2}^{R}(E)\right]$, where $f_{1,2}$ are the Fermi distribution functions of the two contacts.

Starting from the non-equilibrium density matrix $\rho$, via the Mulliken charge analysis \cite{Elstner}, we obtain the non-equilibrium fluctuation in the number of electrons of the $i_{th}$ atom
\begin{equation} \label{eq:chr}
\Delta q_{i}=\sum_{\mu\in i}\sum_{\nu}Re\left[\rho_{\mu\nu}S_{\mu\nu}\right]-q_{i}^{0},
\end{equation}
where $S_{\mu\nu}$ are the elements of the overlap matrix and $q_{i}^{0}$ are the atomic charge of the neutral atoms.

The non-equilibrium fluctuation in the total charge density is in turn the starting point to obtain the self-consistent electrostatic potential, $V({\bf r})$, by solving a three-dimensional Poisson equation in real space with multigrid algorithm:
\begin{equation}
\label{eq:poisson}
\nabla\cdot\left[\nabla\epsilon ({\bf r})V({\bf r})\right]=-4\pi\sum_{i}\Delta q_{i} n_{i}({\bf r}),
\end{equation}   
where the $n_{i}({\bf r})$ are atom-centered, s-like orbital functions with exponential decay \cite{Elstner}.

The current flowing in the system can also be computed starting from $G^{<}_{\mu\nu}$.
The expression for the coherent contribution to the current just recovers the usual expression obtained in Landauer formulation \cite{Datta}
\begin{equation}
\label{eq:curr}
I_{coh}=\frac{2e}{h}\int^{+\infty}_{-\infty}Tr[\Gamma_{1}G^{R}\Gamma_{2}G^{A}](f_{1}-f_{2})dE,
\end{equation}
where $Tr[\Gamma_{1}G^{R}\Gamma_{2}G^{A}]$ is the transmission function computed for a given energy value.


\section{Bulk-switched CNTFETs: the model}


\begin{figure}
\centering
\includegraphics[width=9.0cm,angle=0]{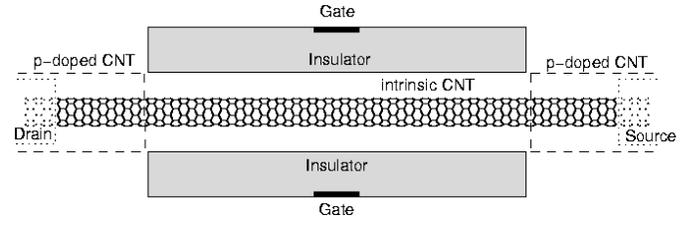}
\caption{Schematic cross section of the coaxially gated CNTFET we have simulated.
\label{fig:system}}
\end{figure}

The system we have simulated is shown in Fig. \ref{fig:system}.
It consists of an infinite, semiconducting, CNT (7,0), having a diameter of $5.52$ {\AA} and an energy gap $E_{g}=1.3$ eV. 
The nanotube is coaxially gated by a metallic cylindric contact centered in the middle point of the channel. 
The gate length is $1.2$ nm.
To simulate the metallic gate electrostatics, the Poisson equation is solved with Dirichlet boundary conditions imposed on the cylindrical surface defining the gate itself.
An insulator layer with dielectric constant $\epsilon_{r}=3.9$ and a thickness of $0.8$ $nm$ surrounds the metallic cylinder.

In order to achieve a bulk-switched device, the electrostatic coupling between the gate and the contacts SBs has to be made negligible. 
This guarantees that the SBs are no longer modulated by the gate field; their role in the device operation simply reduces to fix the carrier injection from the contacts to the channel.
Since details on the metal-nanotube interface are not necessary in order to describe bulk-switching mechanisms, a bulk-switched device can be simulated considering just the nanotube structure without metallic contacts, once a suitable carrier injection modeling has been provided.
 
To apply the NEGF formalism, the nanotube has been divided into three parts: a central region, representing the effective device for which all the non-equilibrium quantities are computed, and two semi-infinite external portions acting as source and drain reservoirs. 
To simulate charge injection in the intrinsic channel, we simply dope the two ending portions of the nanotube. 
The doped contacts provide charge carriers in the CNT valence or conduction band. 
In this manner we exactly recover the experimental situation obtained by modulating the contacts SBs by means of either a back-gate contact or an induced charge transfer with dopant molecules \cite{Chen}.     
To dope the nanotube, we vary the number of valence electrons per carbon atoms. In this doping model any added extra charge is uniformly distributed over the whole doped region, leaving the electronic band structure unaffected \cite{Tersoff}. 
The only effect produced consists of a variation of the carrier density and a consequent shifting of the Fermi level inside the CNT conduction or valence subbands (for $n$ or $p$ doping respectively).
We note that carrier density variation via charge transfer mechanisms is actually the doping process experimentally observed for CNTs.

Calculations have been performed by using an optimized, $sp^{3}$ minimal basis set. 
In order to reduce computational time, we have omitted the $sp^{2}$-hybridization part of the Hamiltonian, retaining only the $p_{z}$-orbital part. 
This is sufficient to accurately describe the first conduction and valence bands of the CNT (7,0). 
Consistency checks between calculations with a reduced, $p_{z}$, and an extended $sp^{3}$ parametrization have been performed to confirm the accuracy of our results.


\section{Screening properties of a gated CNT}
\label{sect:CNT_screen}      

 
Differently from SB-CNT transistors, which operate as contact-dominated switching devices, bulk-modulated CNTFETs operate with mechanisms very similar to standard solid-state nano-MOSFETs. 
Both in CNTFETs and in nano-MOSFETs the source-drain current is modulated by varying the barrier height for carrier propagation in the middle of the channel. 

Evidently, the gate capability of modulating the barrier is determined by the screening properties of the channel itself. In particular, modulation depends on the channel capability to screen the gate field, that is, on the amount and the distribution of charge that can be locally induced in the channel itself by the applied gate bias.
A channel with very good screening properties, an ideal metallic channel, would completely screen the gate potential on its surface; consequently the gate would not be able to modulate the electrostatic potential on the channel.
On the contrary, the gate filed penetrates deep inside a channel with low screening properties. In this case the gate bias would result in a finite electrostatic potential inside the channel, allowing a good gate control of the potential barrier for carrier transmission.  

A quasi-one-dimensional system like a nanotube behaves differently from both the two situations we have just described.
Owing to their very small, one-dimensional density of states (DOS), nanotubes are generally not able to completely screen the gate field in the small extension of the electron gas associated to the graphene sheet.
In other words, the nanotube cannot accumulate all the charge needed for total screening.
In this situation, the channel charge response for a given gate bias can no longer be described by using only a geometrical capacitance, which is derived in classical electrostatics assuming that external fields are completely screened inside the conductors.
On the contrary, a correction should be brought to the geometric capacitance to account for the correct value of the induced charge. This correction is obtained by means of the so called quantum capacitance \cite{Luryi}.
The total capacitance of the gated nanotube system, $C_{G}$, is given by the series connection of the geometric and the quantum capacitances \cite{Buttiker}, respectively $C_{ins}$ and $C_{Q}$, as 
\begin{equation}
\label{eq:capac}
\frac{1}{C_{G}}=\frac{1}{C_{Q}}+\frac{1}{C_{ins}}.
\end{equation} 
Note that the geometric capacitance of the system is just the capacitance of the insulator layer which separates the gate and the nanotube, while the quantum capacitance takes into account the effective charge which can be accumulated by the electron gas associated to the wrapped graphene sheet.
A correct evaluation of the quantum capacitance is fundamental in order to correctly describe the charge response (and consequently the barrier height modulation properties) of a bulk-modulated CNTFET.  
The approach we use in this work naturally includes the treatment of the quantum capacitance, since the charge induced on the nanotube is computed self-consistently with the gate electrostatics. 
Furthermore, by using an atomistic, DFT-based approach, the computed charge takes into account the DOS of the 1D system and retains, at least within a mean-field approximation, some of the electron-electron exchange and correlation effects on the nanotube screening properties.
The importance of many-body corrections to the quantum capacitance of a gated nanotube has been recently pointed out in Ref. \cite{Latessa} and \cite{Fogler}. 
In large diameter tubes, where the contribution of the exchange interaction to the total energy is negligible, the quantum capacitance can be estimated simply from the DOS as $C_{Q_{DOS}}=e^{2}\rho_{0}\left(\epsilon_{F}\right)L$. 
However, the physical behavior drastically changes for small diameter tubes. 

This can be easily seen starting from the most general expression for the quantum capacitance of the nanotube electron gas \cite{Latessa,Fogler},
\begin{equation}
\label{eq:CQ_TDOS}
C_{Q}=e^{2}\frac{dn}{d\mu}L=e^{2}\chi(\epsilon_{F})L,
\end{equation}
where $\mu$, $n$ and $L$ are respectively the chemical potential, the carrier density and the length of the electronic system. 
The quantity $\chi=\rho_{0}K/K_{0}$ is the thermodynamic density-of-states (TDOS) of the nanotube, defined starting from the DOS $\rho_{0}$ of the non-interacting electronic system, and the ratio between the interacting ($K$) and the non-interacting ($K_{0}$) thermodynamic compressibilities.
If electron-electron interactions are neglected ($K=K_{0}$), the quantum capacitance simply accounts for the charge accumulation allowed by the nanotube non-interacting DOS, resulting $C_{Q_{DOS}}=e^{2}\rho_{0}(\epsilon_{F})L$. 
This low, quasi-one-dimensional DOS can only give rise to a partial screening situation.
On the contrary, the thermodynamic DOS, $\chi$, generalize the standard expression used for the quantum capacitance to the case in which electron-electron interactions effects are included at different levels of approximation.
Due to the effect of interactions on the compressibility, $\chi$ can be quite different with respect to $\rho_{0}$.
Indeed, when the kinetic energy is small and the exchange interaction becomes the prevalent contribution to the total energy of the nanotube electronic system, the TDOS can even assume negative values, leading to a negative quantum capacitance.
From a physical point of view, this means that the nanotube, owing to the predominance of the attractive exchange interaction among electrons, can accumulate even more charge than what strictly needed to totally screen the gate field, giving rise to a small over-screening of the gate potential in its interior.
As a matter of fact, when $C_{Q}<0$, from Eq. (\ref{eq:capac}) we obtain $C_{G}>C_{ins}$, and therefore a total induced charge greater than in the classical case.
A negative sign of the TDOS is indeed a property of any strongly correlated electronic system \cite{Dolgov}, and practically manifests itself in charge response effects if the system is a low dimensional one.

\begin{figure}
\centering
\includegraphics[width=5.5cm,angle=270]{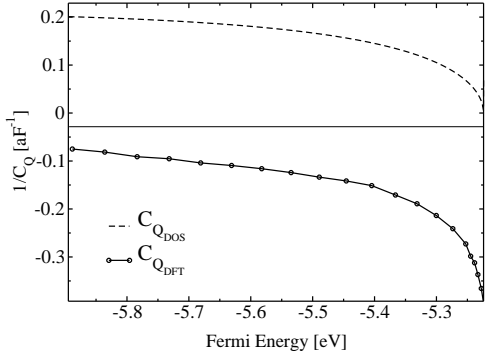}
\caption{Computed inverse quantum capacitance of the coaxially gated CNT (7,0) as a function of the Fermi level inside the first valence subband. Solid line refers to DFT calculations. Dashed line refers to $C_{Q}=e^{2}\rho_{0}(E_{F})$.\label{fig:capac}}
\end{figure}

It is worth mentioning that a negative sign of the thermodynamic compressibility was firstly measured for a two-dimensional electron gas (2DEG) realized with GaAs quantum wells \cite{Eisenstein}.   
In the case of nanotubes, the effects of the exchange interaction has only been predicted theoretically \cite{Latessa,Fogler}, but is expected to be especially appreciable in small diameter tubes with low carrier densities \cite{Latessa_JCE_I}.
In order to explore these effects we have used a lightly-doped small diameter CNT (7,0) for our calculations.

In Fig. \ref{fig:capac} we report the computed inverse quantum capacitance of the $p$-doped CNT (7,0) as a function of the Fermi energy inside the first CNT valence subband.
Each value of the Fermi energy univocally corresponds to a given $p$-doping fraction.
These results have been obtained following the methodology presented in Ref. \cite{Latessa}, and refer to an insulator capacitance $C_{ins}=0.28$ aF.
Solid line refers to the DFT atomistic computation, while dashed line is representative of the DOS-limited, non-interacting result.
The nanotube over-screens the external gate field in the whole range of holes densities we have explored, and no positive values of the quantum capacitance have been obtained.

Note that the smallest absolute value of the quantum capacitance, corresponding to the largest over-screening of the gate potential, has been obtained for the lowest $p$-doping fraction.
In fact, the smallest hole density in valence band corresponds to the smallest contribution of the kinetic energy on the exchange interaction, owing to the fact that carrier velocity vanishes as the subband edge is approached.
The exchange interaction predominance on the kinetic contribution progressively diminishes as the valence subband is filled by holes, leading to larger absolute values for $C_{Q}$ and weakening the over-screening effect.

Finally, it is worth spending some words about the thermodynamic stability of the system.
If we consider the nanotube system as itself, without including the external surface charge on the metallic gate contact, negative values of the thermodynamic compressibility seems to make the system unstable under the spontaneous appearance of an electric field or a charge fluctuation \cite{Dolgov}.
Indeed, the inclusion of the external surface charge on the gate, represented here by the inclusion of the insulator capacitance, makes the total gate capacitance of the whole system a positive quantity.   
In other words, for the total system, nanotube plus gate contact, the negative TDOS does not imply any instability, but leads instead to a small overs-screening of the external field.
A practical stability criterion for our system can be therefore formulated as the requirement for a positive $C_{G}$. 
From Eq. (\ref{eq:capac}), it is straightforward to be seen that the stability condition is therefore ensured if $\left|C_{Q}\right|\ge C_{ins}$, a condition always fulfilled in our calculations.


\section{CNTFET output characteristics}
\label{sect:IVout}


\begin{figure}[b]
\centering
\includegraphics[width=5.8cm,angle=270]{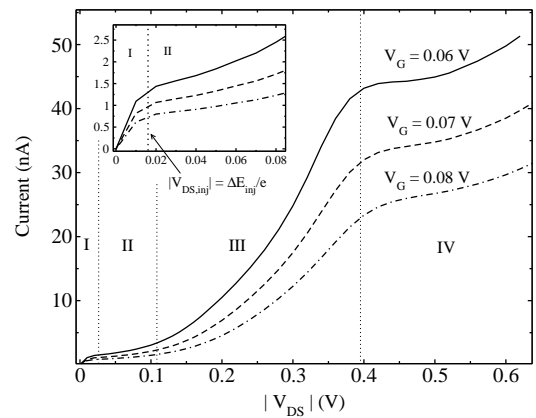}
\caption{Computed output characteristics of the $p-i-p$ CNTFET. Different curves correspond to different gate potential. The inset shows details of the first two operative regimes of the device, the linear regime and the carrier injection saturation (see the text for details).
\label{fig:IVDS}}
\end{figure}

\begin{figure}[t]
\centering
\includegraphics[width=6.0cm,angle=270]{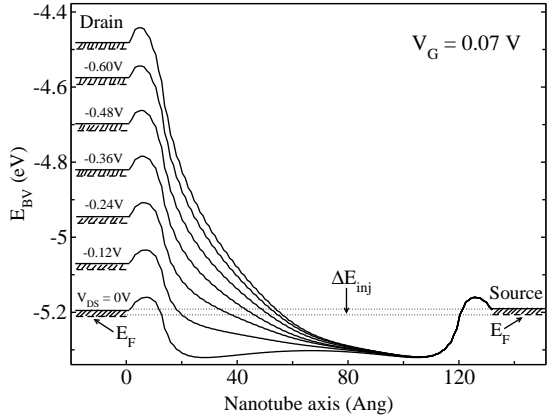}
\caption{Computed valence band profiles for the $p-i-p$ structure as a function of the voltage applied to drain contact. The gate voltage is fixed to $V_{G}=0.07$.
\label{fig:Bands}}
\end{figure}

In Fig. \ref{fig:IVDS} we show the calculated output characteristics for a $p-i-p$ CNTFET. 
Clearly, we can identify at least four distinct regions for current behavior, which will be discussed in detail later. 

The intrinsic channel of the simulated nanotube is $10.26$ nm long. Source and drain contacts are $p$-doped with a carrier concentration of $2.63\cdot 10^{6}$ $cm^{-1}$, corresponding to a doping fraction $f=0.004$ holes per carbon atom. 
In realistic devices, this carrier concentration can be easily obtained by an electrostatic doping, controlled by means of a back gate contact \cite{Lin_condMat}.  
This doping causes the nanotube Fermi level to be shifted inside the first valence subband, at $\sim16$ meV below the subband edge. 
Charge transfer at the junction between the degenerate $p$-type and the intrinsic portion of nanotube generates a barrier for holes transmission in valence band. 
The computed valence band profile is the lowest curve shown in Fig. \ref{fig:Bands}, corresponding to the situation in which no drain bias is applied.

\subsection{Band profile and boundary conditions}

\begin{figure}[b]
\centering
\includegraphics[width=6.0cm,angle=270]{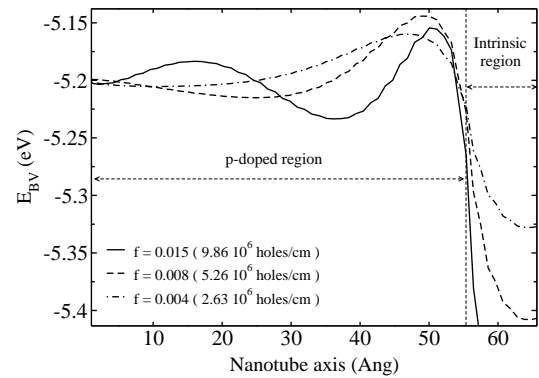}
\caption{Details of the computed band profile at the interface between the doped and intrinsic portions of the nanotube. Friedel oscillation are clearly shown. Different curves have been obtained by varying the doping fraction of the contacts.
\label{fig:oscill}}
\end{figure}

Results in Fig. \ref{fig:Bands} clearly show that the electrostatic potential does not reach smoothly a constant value as the unperturbed source and drain contacts are approached. On the contrary, it oscillates around a constant value which is reached only when oscillations are dumped far from the interface between the doped and intrinsic portion of the nanotube.
A more detailed representation of the band profile at the interface between the intrinsic and doped region is given in Fig. \ref{fig:oscill}, where the computation has been performed by using a more extended contact doped region and for various $p$-doping fractions. 
These oscillations in the electrostatic potential are just the Friedel oscillations associated to the screening charge distribution, which appear if electron-electron interactions effects are considered beyond the simple Thomas-Fermi model \cite{Fetter}.
These many-body corrections to screening are obviously considered in our DFT approach.
The oscillation period of the screened potential (Fig. \ref{fig:oscill}) is smaller correspondingly to higher $p$-doping fractions, i.e., for larger Fermi $k$ vectors of the electronic system, $k_{F}$, in agreement with the Friedel oscillations period, known to be proportional to $1/k_{F}$ \cite{Calmels95}.

We note that in the band profile shown in Fig. \ref{fig:Bands} the oscillations in the doped regions are not completely dumped before the contact-device interface.
Nevertheless, we have found that this does not affect the band profile inside the intrinsic channel, which practically does not change appreciably if the doped regions are further extended.

In our case, very long doped regions are not needed to ensure the correct computation of the non-equilibrium band profile.
The reason for this is quite subtle, and has been extensively discussed within the simulation techniques used for nano-MOSFETs (see, for instance, Ref. \cite{Venugopal}).
Here we only observe that, in our device, a sufficiently small holes tunneling probability maintains a near-equilibrium charge distribution in the source and drain $p$-doped region.
Therefore, the electrostatic potential in these regions has not to be shifted as a function of the applied drain bias in order to account for charge neutrality.
For this reason, we do not strictly need to use the correct BCs for the potential, neither by using sufficiently long $p$-doped regions, nor by employing Neumann BCs which leave the potential completely floating to be adapted to the correct non-equilibrium charge accumulation.  
We can obtain the same correct results for transport by using Dirichlet BCs at a small distance from the doped-intrinsic interface.

\subsection{Current saturation mechanisms}

The different curves shown in Fig. \ref{fig:IVDS} refer to different gate biases. 
Positive potentials applied to the gate contact increase the barrier height for holes transmission, hence reducing the current for a given drain voltage. 

The physical mechanisms governing the current dependence on the applied source-drain bias can be understood referring to the band profiles shown in Fig. \ref{fig:Bands}. These band profiles have been obtained, for a fixed gate potential, by varying the drain-source applied bias, $V_{DS}$, in a range of negative values (as required for a $p$-type conduction device).

Following the result given in Eq. \ref{eq:curr}, the current is determined by the holes transmission probability in the energy interval between the Fermi levels of the source and drain contact.
Although the applied bias continuously shifts the drain Fermi level with respect to the source, the nanotube bandgap reduces the energy window for hole injection to the energy interval $\Delta E_{inj}$, between the source valence band edge and the source Fermi level (see Fig. \ref{fig:Bands}).
In other words, the maximum holes injection in the channel is fixed by the source doping rate.
Following Fig. \ref{fig:IVDS}, we can detail at least four fundamental, distinct mechanisms which govern the current flowing across the channel.

The first (regime I in Fig. \ref{fig:IVDS}) is characterized by a linear increase of the current as a function of the applied bias, simply due to the increasing of the energy window for carrier injection between the source and drain Fermi levels.
The linear behavior is predicted by the Landauer formula in Eq. \ref{eq:curr}, assuming that the transmission probability does not depend on the (small) applied bias. 

\begin{figure}
\centering
\includegraphics[width=5.8cm,angle=270]{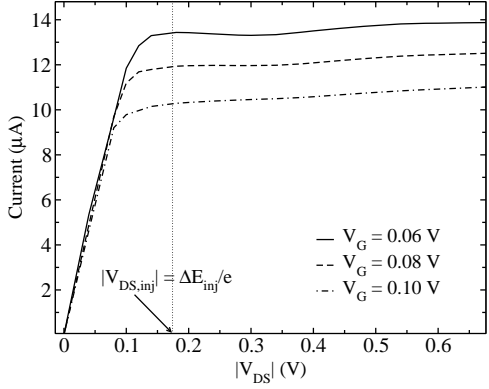}
\caption{Output characteristics for the $p-i-p$ structure with a doping fraction of $0.004$ holes per carbon atom and a gate bias of $0.07$ V. The lowest curve is just the output characteristic already shown in Fig. \ref{fig:IVDS}. The other curves have been obtained for the same band profile, but increasing the holes injection interval, reducing the current sensitivity to the drain-induced barrier variation. 
\label{fig:IVDS_Inj}}
\end{figure}

As the maximum energy window for carrier injection is reached, corresponding to the bias $|V_{DS,inj}|=\Delta E_{inj}/e$, a second regime is entered (regime II in Fig. \ref{fig:IVDS}).
The transition from the first to the second operative regimes can be appreciated in the inset of Fig. \ref{fig:IVDS}, where the output characteristics are detailed in a small bias range. The drain bias, $V_{DS,inj}$, corresponding to a shift $\Delta E_{inj}$ between the source and drain Fermi levels, is indicated.
In standard nano-MOSFET theory, the second regime effectively corresponds to current saturation, since the carrier transmission probability in the interval $\Delta E_{inj}$ is assumed to be no longer modulated by further increasing the applied bias $V_{DS}$.
A ``well-tempered'' device fulfills these requirements.
The output characteristics shown in Fig. \ref{fig:IVDS} reveal that this is not the case of our short-channel CNTFET.   
The drain-induced modulation of transmission probability within the injection interval leads to a relatively small increment of current as long as the drain bias locally lowers the barrier close to the drain contact itself; this is for instance the case of the lowest band profiles in Fig. \ref{fig:Bands}.
A ``partial'' saturation can still be identified in this case.
 
When the barrier starts to be also narrowed inside the injection interval, a third regime is entered (regime III in Fig. \ref{fig:IVDS}), corresponding to a much more rapid increase in current.
The reason for the non-saturation can be understood looking at the band profile in Fig. \ref{fig:Bands}. 
As $V_{DS}$ increases, the effective barrier length reduces, causing transmission to increase.
It should be kept in mind that the device is now operating in such a regime that current is entirely due to tunneling through the potential barrier.
  
The fourth regime is entered for large $V_{DS}$, where a partial current saturation appears again in the output characteristics (regime IV in Fig. \ref{fig:IVDS}).
After a regime in which the current continuously increases with $V_{DS}$, the differential conductance of the channel ($dI_{DS}/dV_{DS}$) abruptly diminishes giving rise to a plateau in the output characteristics.
The reason for this behavior can be explained referring again to the band profiles in Fig. \ref{fig:Bands}. 
Here it can be seen that the drain-induced narrowing of the barrier eventually saturates for large $V_{DS}$, giving rise to a progressive reduction of the differential conductance of the channel. 
We can identify this new regime as an electrostatics-dominated current saturation to distinguish it from the well known injection-limited saturation mechanism, which instead governs the behavior of standard nano-MOSFET.

Indeed, a further injection contribution, and therefore a fifth regime, can appear for very large biases, when a Zener band-to-band tunneling is reached. 
However, the large bandgap nanotube we have used for our calculations ($E_{g}=1.3$ $eV$) prevents this regime to be reached in the explored bias range.

It should be noted, from Fig. \ref{fig:Bands}, that the top of the barrier for hole transmission is well pinned as the drain bias is increased.
Therefore, we cannot speak of drain-induced barrier lowering (DIBL) in this particular case.
The role of the doping profile in generating a DIBL-governed behavior will be discussed in section \ref{sect:screen}.
Here it is important to note that, although the top of the barrier is pinned by the gate bias, the drain-induced change in the valence band profile is the fundamental reason for the simulated behavior of the output characteristics.

In purely ballistic devices, it is just the barrier height which determines the device behavior.
On the contrary, in tunneling regimes, also the barrier width and shape are relevant in determining the current.
In the $p-i-p$ structure we have simulated, the contact Fermi level lies closely to the valence subband edge, well below the top of the potential barrier for hole transmission (see Fig. \ref{fig:Bands}).
This is not totally unexpected for a small-diameter nanotube junction, where charge transfer occurring at the doped-intrinsic interface is distributed over an exiguous number of carbon atoms, owing to the marked quasi-one dimensionality of the system.
A large shift in the atomic energy can therefore results also from a relatively small charge transfer, causing a very high potential barrier with respect to carrier injection level.
This deeply differentiates the behavior of the device we are studying here with respect to the ballistic transport theory developed for silicon nano-MOSFET \cite{Natori}.
The strong dependence of the saturation current on the drain bias we have so far observed, is just caused by the exponential dependence of the tunneling current on the width of the barrier.   
We can refer to the mechanism responsible for the non-saturation of the current as drain-induced barrier thinning (DIBT).

As a larger injection interval is used, introducing also a ballistic current contribution, the dependence of the saturation current on the drain bias is reduced.
This can be observed in Fig. \ref{fig:IVDS_Inj}, where we show the output characteristics of the CNTFET obtained by evaluating the current with an injection interval of $0.186$ eV, which simulates a contact Fermi level above the top of the potential barrier for holes transmission.
Clearly, as a ballistic contribution is introduced to the carrier injection, the dependence of the injection-saturated current on the drain bias is drastically reduced.
In fact, all the three curves show a practically linear regime corresponding to the pure ballistic transmission, followed only by the regime of current injection saturation, closely resembling the output characteristics of a standard silicon nano-MOSFET. 
It is worth noting that the gate capability of pinning the top of the potential barrier ensures also in this case different saturation currents for different gate applied biases, allowing for an effective modulation of the output current in the saturation regime.

Finally, it should be noted that the short length of the channel enhances the sensitivity of the tunneling current to drain-induced barrier modulation.
Longer nanotubes are therefore expected to show a considerably lesser sensitivity of the saturation current to the drain bias.

\begin{figure}
\centering
\includegraphics[width=7.0cm,angle=0]{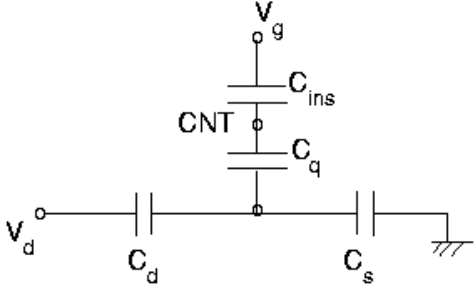}
\caption{Capacitive circuital model for the coaxially gated CNTFET.
\label{fig:circ}}
\end{figure}

\subsection{Screening effects on transport}
\label{sect:screen}

In this section we analyse the screening properties of the nanotube in relation to the field-effect device functionality.
Screening is directly tied to the gate capability of pinning the top of the potential barrier, and therefore controls the DIBL mechanism.   
Calculations we present are also intended to explore the possibility of a well tempered device design, ensuring an almost ideal gate control in CNTFETs.

Substantial differences can arise with respect to bulk nano-devices due to the peculiar one-dimensional screening properties of the nanotube, as discussed in Section \ref{sect:CNT_screen}. 
In particular, we emphasize here the role of the quantum capacitance in determining the behavior of the transistor output characteristics.
In principle, quantum capacitance should be included also in the models describing transport in nano-MOSFETs, since the small dimensions of the device could introduce limitations in the screening response of the channel.     
However, the effects of the quantum capacitance become remarkable only when they give a large contribution to the total gate capacitance of the system. 
This means that an effective quantum capacitance regime is generated only if $C_{Q}\ll C_{ins}$. This regime is hardly reached in nano-MOSFETs, even for very small insulator thicknesses giving a large $C_{ins}$. 
On the contrary, the quasi-one dimensional DOS of a nanotube is naturally associated to a very small quantum capacitance which can dominate the screening properties of the channel.

In order to analyse the effect of screening on the output characteristics, we refer to the circuital model shown in Fig. \ref{fig:circ}. 
The total gate capacitance $C_{G}$, given by $1/C_{G}=1/C_{ins}+1/C_{Q}$, accounts for the charge induced on an infinite, uniformly doped nanotube when a voltage $\delta V_{G}$ is applied to the gate contact.
$C_{Q}$ accounts for the effects on charge response tied to both the low-dimensionality of the system and to the many-body nature of the screening problem.
$C_{S}$ and $C_{D}$ take into account the charge induction of the source and drain $p-i$ junctions.  

Solving the circuit model of Fig. \ref{fig:circ}, the average electrostatic potential $\delta V_{CNT}$ induced on the nanotube can be obtained as 
\begin{equation}\label{eq:pot}
\delta V_{CNT}=\alpha_{G}\delta V_{G}+\alpha_{D}\delta V_{D},
\end{equation} 
where
\begin{equation} \label{eq:alphaG}
\frac{1}{\alpha_{G}}=1+\frac{C_{Q}'}{C_{ins}}
\end{equation}
and
\begin{equation} \label{eq:alphaD}
\frac{1}{\alpha_{D}}=2\left(1+\frac{C_{ins}}{C_{Q}'}\right)=\frac{2}{1-\alpha_{G}}.
\end{equation}
The quantities $\alpha_{G}$ and $\alpha_{D}$ measure the gate and drain control over the energy barrier.
Solving the elementary circuit equations we obtain 
\begin{equation}
\label{eq:CQ'}
C_{Q}'=\frac{2C_{D}C_{Q}}{C_{Q}+2C_{D}}.
\end{equation}
Note that the expression for both $\alpha_{G}$ and $\alpha_{D}$, as well as the definition of $C_{Q}'$, have been obtained under the assumption that $C_{D}=C_{S}$.
This assumption is obviously valid only in the case of a symmetric doping profile, just as in our case. 

\begin{figure}[b]
\centering
\includegraphics[width=7.0cm,angle=270]{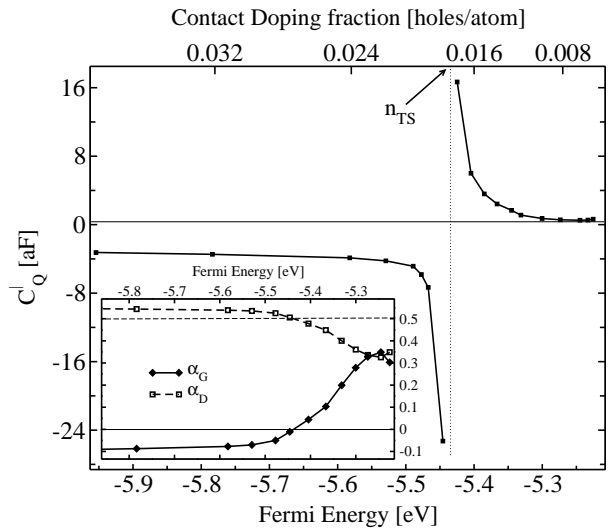}
\caption{Quantum capacitance $C_{Q}'$ of the $p-i-p$ structure as a function of the Fermi level of the system. Different values correspond to different contact doping fractions. The inset shows the corresponding gate and drain control factors, $\alpha_{G}$ and $\alpha_{D}$. An insulator capacitance $C_{ins}=0.28$ aF has been used for this calculation.
\label{fig:CQDS}}
\end{figure}
 
The capacitance $C_{Q}'$ is just given by the series connection of $C_{Q}$ with the total drain-source capacitance $C_{D}+C_{S}$ (see the circuital model in Fig. \ref{fig:circ}).
It can be therefore intended as a quantum capacitance defined for a non-uniformly doped nanotube. In fact, it corrects the standard quantum capacitance of the channel by including the effect of the source and drain spatial charge regions in the nanotube charge response.
The numerical results for $C_{Q}'$, shown in Fig. \ref{fig:CQDS}, have to be compared with those obtained for the uniformly doped channel, shown in Fig. \ref{fig:capac}.
The effects of the non-uniform doping profile are clearly important.
One can immediately note that the $p-i-p$ doping profile, in contrast to the uniformly doped nanotube, results into an over-screening behavior only for large $p$-dopings.
For a very low $p$-doping value, $C_{Q}'$ is a positive quantity and the nanotube partially screens the gate field.

The knowledge of $C_{Q}'$ is sufficient in order to directly obtain the gate and drain control factors of the transistor from Eq. \ref{eq:alphaG} and \ref{eq:alphaD}.
It can be straightforwardly seen from Eq. (\ref{eq:alphaG}) that the best gate control corresponds to the condition $C_{Q}'<< C_{ins}$, which gives an $\alpha_{G}$ very close to unity, leading to an optimal  gate capability of pinning the electrostatic potential on the nanotube.
At the same time, when $C_{Q}'<< C_{ins}$, a very small fraction $\alpha_{D}$ of the drain applied bias will affect the potential profile on the CNT (see Eq. \ref{eq:alphaD}).  
This practically means that the barrier height for carrier transmission will be fixed by the gate, diminishing the influence of the drain bias.
The computed gate and drain control factors, $\alpha_{G}$ and $\alpha_{D}$, are plotted in the inset of Fig. \ref{fig:CQDS} as a function of the Fermi energy of the system.
The best behavior for our device will be obtained for a small contact doping fraction, $f=0.004$ holes per carbon atoms ($2.63\cdot 10^{6} cm^{-1}$), for which a gate control factor $\alpha_{G} \simeq 0.35$ and a drain control factor $\alpha_{D} \simeq 0.32$ have been found. 
A quantum capacitance $C_{Q}'=0.52$ aF and an insulator capacitance $C_{ins}=0.28$ aF have been obtained for this case.
The output characteristics shown in Fig. \ref{fig:IVDS} refer to this situation.
They therefore represent the best behavior which can be obtained in terms of drain-induced barrier modulation for the given device geometry we have chosen, i.e., for the given insulator capacitance.
Note that the relatively small value of the gate control factor obtained for this optimal case  is indeed sufficient to pin the top of the barrier height closely to the source contact (see the band profiles in Fig. \ref{fig:Bands}).

It should be noted, from the inset in Fig. \ref{fig:CQDS}, that $\alpha_{G}$ can also assume negative values.
This unconventional situation is associated to the fact that the CNT (7,0) is able to over-screen the gate field.
In this case, in fact, a positive variation of the applied gate bias produces a negative change in the electrostatic potential on the nanotube.
Note that a large, but negative $\alpha_{G}$ is not able to generate an optimal gate control capability since it corresponds to a large, positive drain control factor $\alpha_{D}$, which becomes larger than $0.5$ when the over-screening regime is entered.
The results shown in the inset of Fig. \ref{fig:CQDS} clearly show what have just stated, as the operation in the negative quantum capacitance regime can only make the performances of the nanotube device worse.

\begin{figure}
\centering
\includegraphics[width=6.3cm,angle=270]{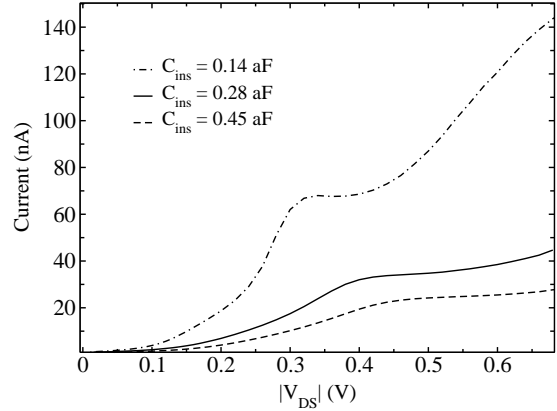}
\caption{Output characteristics of the simulated CNTFET. A gate bias $\delta V_{G}=0.07$ $V$ and a contact doping fraction $f=0.004$ holes per carbon atom ($2.63\cdot 10^{6} cm^{-1}$) have been used. 
Solid line is just the output characteristic shown in Fig. \ref{fig:IVDS}, corresponding to $C_{ins}=0.28$ aF. Dashed and dot-dashed lines correspond to $C_{ins}=0.14$ aF and $C_{ins}=0.45$ aF.
The gate and drain control factors have changed from $\alpha_{G}=0.21$ and $\alpha_{D}=0.39$ in the former case, to $\alpha_{G}=0.46$ and $\alpha_{D}=0.27$ in the latter case.  
\label{fig:IVins}}
\end{figure}

\begin{figure}
\centering
\includegraphics[width=6.3cm,angle=270]{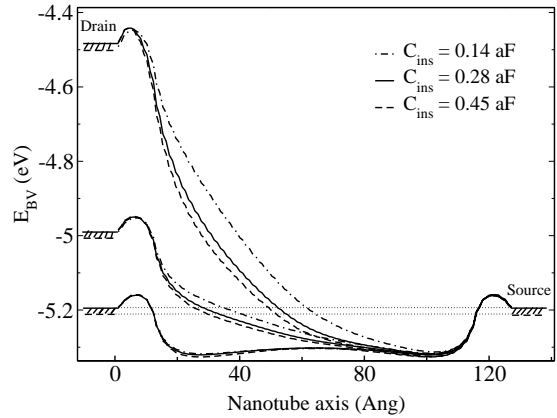}
\caption{Computed valence band profiles corresponding to the output characteristics in Fig. \ref{fig:IVins}. Different profiles corresponding to the same value of the drain bias have been obtained for different insulator capacitances. 
\label{fig:Bandins}}
\end{figure}

\subsection{Toward a ``well-tempered'' design: gate geometry and contact doping}

The most simple approach to control the gate modulation capability for a given doping profile of the nanotube, consists of changing the geometric parameters of the system, such as the gate insulator capacitance. 
In Fig. \ref{fig:IVins} we show the effect of variations of $C_{ins}$ on the device characteristics.
Clearly, a higher insulator capacitance improves the gate control while reducing the influence of the drain bias on the potential barrier.
The gain and drain control factors obtained for the different insulator capacitances are listed in the figure.
The valence band profiles corresponding to the three output characteristics of Fig. \ref{fig:IVins} are shown in Fig. \ref{fig:Bandins}.
The computed band profiles confirm that the barrier is much more sensitive to a variation in the drain bias if the insulator capacitance is reduced.
This causes a more rapid increasing of the current with $V_{DS}$, and hence a worse current saturation condition, in both the second and the fourth saturation regimes defined earlier in Fig \ref{fig:IVDS}.

\begin{figure}
\centering
\includegraphics[width=5.8cm,angle=270]{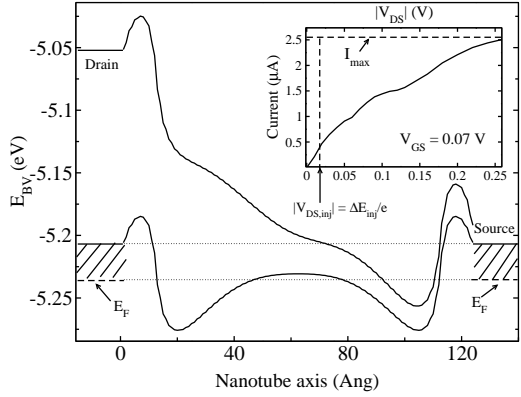}
\caption{Valence band profile and output-characteristic (inset) for a $p^{+}-p-p^{+}$ CNTFET. The device operates in an over-screening regime, which generates a very poor gate control capability. As a matter of fact, no current saturation can be distinguished in the computed output-characteristic.  
\label{fig:Dop_ch}}
\end{figure}

In Fig. \ref{fig:Dop_ch} we show the output characteristics (inset) and the valence band profiles obtained for a $p^{+}-p-p^{+}$ CNTFET in which both the contacts and the channel are $p$-doped, with a doping fraction of $0.004$ and $0.001$ holes per carbon atom respectively.
The channel is no longer intrinsic, but by using a smaller doping with respect to the contacts, the resulting band profile closely resembles that we have obtained for the $p-i-p$ case.
Nevertheless, in this latest case, the holes density inside the channel is much larger than that produced by charge transfer doping in the intrinsic tube. Consequently, the screening properties of the nanotube are strongly changed.
The higher carrier density in the channel causes the nanotube to enter the over-screening regime (see Fig. \ref{fig:CQDS}), and the quantum capacitance of the non-uniform system assumes a negative value.
In particular, we have obtained $C_{Q}'=-4.1$ aF.
In this case, Eq. \ref{eq:alphaG} and \ref{eq:alphaD} predict a very small negative value for the gate control factor and a strong enhancing of the the barrier lowering effect due to the drain bias, owing to a drain control factor larger then $0.5$.
Indeed, for this case, we obtained $\alpha_{G}=-0.073$ and $\alpha_{D}=0.54$.

The resulting very poor gate control is clearly shown by the calculated valence band profiles in Fig. \ref{fig:Dop_ch}.
The top of the barrier for holes transmission is no longer pinned by the gate; on the contrary, it is considerably reduced already for small drain voltages, and rapidly disappear above the valence band edge allowing an almost ballistic holes propagation through the channel.  
The current flowing through the nanotube, shown in the inset of the figure, rapidly increases with the drain bias, and the four different transport regimes previously described are no longer distinguishable in this case. 
As the barrier for holes transmission disappears with $V_{DS}$, the current approaches the value obtained in the ballistic limit (i.e., for a unitary transmission probability) for the given injection energy interval.
This current is obviously the maximum current that can flow across the device (if we do not consider the Zener tunneling contribution occurring for high drain biases).
The ballistic current obtained for the simulated $p^{+}-p-p^{+}$ structure is $I_{max}=4e/h\cdot\Delta E_{inj}=2.54\cdot 10^{-6}$ A (see Eq. \ref{eq:curr}), where both the spin and band degeneracy of the nanotube have been considered. 

It is worth noting that if the gate is not able to pin the top of the barrier, the current will saturate at $I_{max}$, regardless of the particular voltage applied to the gate contact \cite{Latessa_JCE}.
No gate modulation can therefore be achieved if the transistor is used in the saturation regime.  

The results presented in this section clearly emphasize the important role of the nanotube quantum capacitance in a well-tempered design of a CNTFET.
A good gate control capability can no longer be achieved with a purely geometric design like in standard scaling MOSFET theory.
On the contrary, a deep knowledge of the details of the screening phenomena in the quasi-one-dimensional channel becomes necessary to address the best performances of the device.


\section{CNTFET trans-characteristics}
      
\begin{figure}[b]
\centering
\includegraphics[width=6.8cm,angle=270]{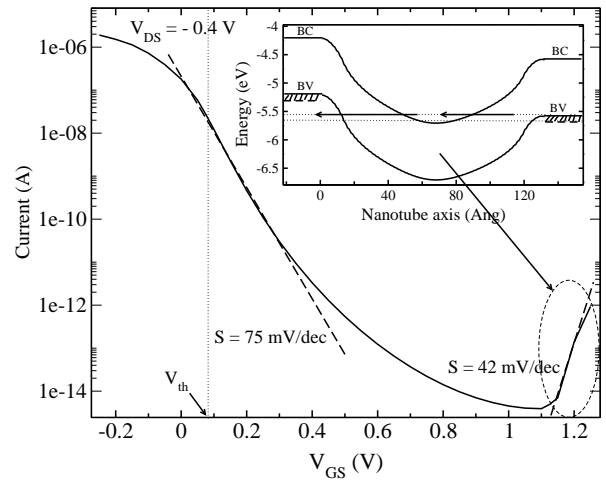}
\caption{Trans-characteristic of a $p-i-p$ CNTFET consisting of the semiconducting CNT (7,0) doped at the contacts with $0.004$ holes per carbon atom. The non-monotonic behavior of the current for high gate voltages is associated to band-to-band tunneling. 
\label{fig:IVGS}}
\end{figure}

In Fig. \ref{fig:IVGS} we show the trans-characteristic obtained for a $p-i-p$ CNTFET.
The simulated device consists of a semiconducting CNT (7,0) doped at the contacts with $0.004$ holes per carbon atom.
The corresponding output characteristics are shown in Fig. \ref{fig:IVGS}. Trans-characteristics have been obtained for a fixed drain bias of $-0.4$ V. 
The device therefore operates in the electrostatic saturation regime.

Our results confirm the exceptional transport characteristics of bulk-modulated CNTFETs.
Firstly, we note that the device shows a perfectly unipolar behavior.
Current is carried only by holes which are injected from the $p$-doped source contact, and is progressively switched-off by increasing the barrier for holes transmission with the gate bias.
The non-monotonic behavior that can be observed in the trans-characteristic when a high gate voltage is applied is associated to band-to-band (BTB) tunneling mechanisms, which allows holes to propagate from the valence subband of source contact into the conduction band in the middle of the channel, and then to be collected into the valence subband of the drain \cite{Appenzeller_BTB} (see the inset in Fig. \ref{fig:IVGS}).
Obviously, a high gate bias is needed to align the conduction band energy levels with the holes injection levels in the source contact. 
A BTB branch has also been recently reported in the experimental trans-characteristics of bulk-modulated CNTFETs \cite{Appenzeller_BTB}.  
The unipolar behavior of this device appears even more interesting if we consider that a very thin gate oxide ($8$ {\AA}) has been used for our simulations.
It has been demonstrated, in fact, that the ambipolar behavior in SB-CNT devices is strongly enhanced if an aggressive vertical scaling of the gate oxide is employed to improve their switching performances \cite{Heinze,Radosav}. 
Thanks to their intrinsic unipolar behavior, vertically scaled CNTFETs accomplish one of the fundamental requirements for a high-performance device which can be used in CMOS-like logic circuits.

\begin{figure}
\centering
\includegraphics[width=7.0cm,angle=270]{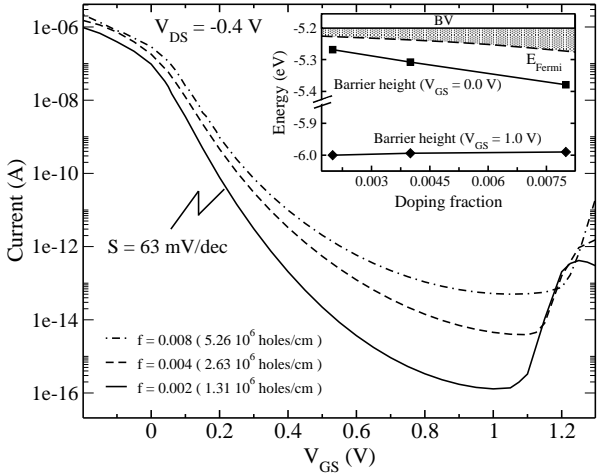}
\caption{Trans-characteristics of the simulated CNTFET. Different curves refer to different contact doping values. The inset shows the dependence of the the system Fermi energy (dashed line) and the height of the potential barrier in valence band (solid lines) on the doping fraction of the nanotube contacts.
Square- and circle-marked lines refer respectively to a gate bias of $0.0$ and $1.0$ V.
\label{fig:IVGS_dop}}
\end{figure}

It is worth mentioning that the ambipolar behavior of SB-CNT transistors drastically increases the $I_{on}/I_{off}$ ratio between the current flowing in the on-state of the device and the minimal current which can be obtained switching-off the channel conductivity.
Fig. \ref{fig:IVDS} shows that the situation is completely different for bulk-modulated CNTFETs.
The computed output-characteristics show a very high $I_{on}/I_{off}$ ratio, $\sim 10^{8}$, which is evidence of the exceptional effectiveness of the bulk-switching mechanism in modulating the current. 

Another  fundamental drawback of SB-nanotube transistors is the high value of the sub-threshold swing parameter, given as $S=dV_{G}/d(logI_{DS})$. 
The computed output characteristic of our bulk-modulated CNTFET shows instead a very good value for the inverse sub-threshold slope, $75$ mV/dec, very close to the measured values reported in Ref. \cite{Appenzeller_BTB,Lin_condMat}.
Note that the BTB branch of the output characteristic in Fig. \ref{fig:IVGS} shows a sub-threshold swing of $42$ mV/dec, which is even lower than the minimum value attainable for any conventional silicon MOSFET at room temperature, which is $K_{B}Tln10/e\simeq 60$ mV/dec.
It should be noted that results we have reported in Fig. \ref{fig:IVGS} refer to $T=0$ K. Indeed, the subthreshold swing in the BTB regime, which is clearly dominated by tunneling currents, is not expected to change by increasing the temperature (see results in the next section and Ref. \cite{Appenzeller_BTB}).
Exceptionally performing, low-power devices could therefore be realized by employing a bulk-modulated CNTFET in the BTB regime even at room temperature.

In Fig. \ref{fig:IVGS_dop} we show the output characteristic of the CNTFET as a function of the contact doping fraction.
By changing the contact doping, both the maximum energy injection interval and the band profile are changed.
In particular, higher doping fractions correspond to larger energy windows for holes injection, shown  by the shaded space between the Fermi level (dashed line) and the valence subband edge in the inset of Fig. \ref{fig:IVGS_dop}. 
At the same time, a larger contact doping cause a larger charge transfer in the channel, which in turn produces a higher barrier for holes transmission.
The inset of Fig. \ref{fig:IVGS_dop} shows the top of the barrier as a function of the doping for $V_{GS}=0.0$ V.
The trans-characteristics depend on the contact doping by the interplay of these two mechanisms.
The results shown in Fig. \ref{fig:IVGS_dop} state that the current increases with the contact doping fraction.
This emphasizes that the increase of carrier injection is more effective than the reduction of tunneling probability.  

Note that the increase of current with doping is not linear with gate bias, but becomes larger and larger as the gate voltage is increased.
This causes the lowest-doping device to show the best performances in term of sub-threshold swing and $I_{on}/I_{off}$ ratio: the steepest subthreshold slope $S\sim 65$ mV/dec and the largest $I_{on}/I_{off}$ ratio $\sim 10^{10}$ have been in fact obtained for the solid line in Fig. \ref{fig:IVGS_dop}, corresponding to the contact doping fraction $f=0.002$ ($1.32\cdot 10^{6} cm^{-1}$).  

The reason for this can be understood referring to the inset of Fig. \ref{fig:IVGS_dop}, where the top of the potential barrier for holes transmission is plotted as a function of the doping fraction for different gate voltages.
At $V_{GS}=1.0$ V (circle-marked line), the top of the barrier practically does not depend on doping, owing to the fact the high gate bias completely depletes the channel of holes for screening.
Therefore, the increase of injection with dopings is not compensated by a corresponding increase of the barrier height. 
The overall effect is then a much larger current increasing with respect to the case at $V_{GS}=0.0$ V.

\subsection{Temperature effects}

All the current calculations reported above have been obtained assuming an electronic temperature of $0$ K.
Nevertheless, we have previously compared the computed value of the subthreshold swing in the BTB regime with the theoretical minimum obtained for standard silicon MOSFET at room temperature.  
This is justified only if the BTB subthreshold slope is not expected to vary with temperature. 

Two main mechanisms can cause the current to change with temperature in our simulations: (i) the temperature dependence of the Fermi distribution in the doped nanotube contacts defining the hole injection, and (ii) the different occupation of the scattering states of the system which makes the charge induced in the channel a temperature dependent quantity.

\begin{figure}
\centering
\includegraphics[width=7.0cm,angle=270]{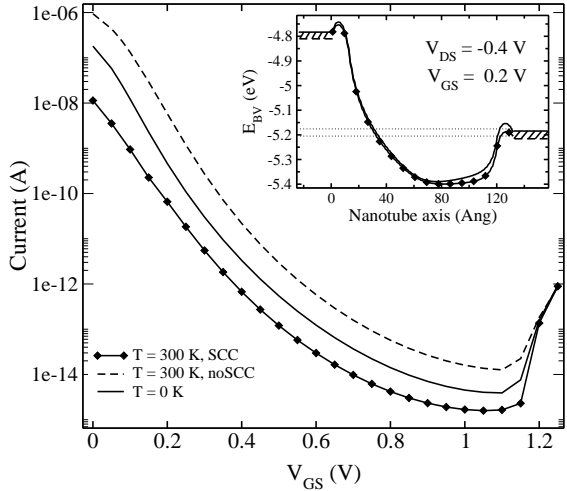}
\caption{Temperature dependence of the subthreshold characteristics and valence band profile (inset). Calculations have been performed for the $p-i-p$ device with a contact doping fraction of $0.004$ holes per carbon atom, at a drain bias of $-0.40$ V. Dashed line represents only the effect of temperature on the Fermi distributions within the contacts. It has been obtained without considering the temperature-induced change in self-consistent charge accumulated on the nanotube, which is instead included for the solid lines with markers.
\label{fig:IVGS_temp}}
\end{figure}

When only the first mechanism is considered, the current simply changes due to the dependence of the tunneling probability with energy. 
In our case, high temperatures favor holes propagating at high transmission energy channels near the top of the barrier, causing a net increase in the current.
The dashed line in Fig. \ref{fig:IVGS_temp} shows the current obtained for this case.

When the temperature is included self-consistently in the non-equilibrium calculation, the behavior of the current is more complex.
As the temperature increases, the overall increasing in the tunneling probability causes in turn a larger accumulation of holes in the nanotube channel. 
Consequently, a lower number of electrons is available to locally screen the positive voltage applied to the gate contact, causing in turn an increasing of the potential barrier for hole transmission with respect to the case for $T=0$ K.
In other words, a finite temperature reduces the channel capability of locally screening the gate field, and a larger fraction of the gate bias drops on the nanotube increasing the potential barrier.
A fully self-consistent computation of the valence band profile and current at room temperature is shown in Fig. \ref{fig:IVGS_temp}.
Clearly, the combined effect of the thermal assisted tunneling and the increasing of the potential barrier results in a reduction of the tunneling current with respect to $T=0$ K.
Since our lightly-doped nanotube is practically emptied of mobile carriers at high gate biases, the potential barrier is not influenced by the temperature-dependent charge accumulation in the BTB regime. 
This causes the BTB-tunneling current to be insensitive to temperature variations, as clearly shown by Fig. \ref{fig:IVGS_temp}. 

It should be noted that the effect of temperature so far discussed is associated only to the energy re-distribution of injected holes inside the doped nanotube contacts according to the Fermi distribution function.
In a realistic device, in which SBs are present at the metal-semiconductor contacts interface, a further contribution of temperature comes from the increasing of the hole density inside the doped nanotube regions. 
In fact, these regions are effectively doped by carriers tunneling through the SBs.
This would generate a self-consistent increasing of the effective doping rate of the nanotube contacts with temperature \cite{Appenzeller_Therm}, which would produce, according to results in Fig. \ref{fig:IVGS_dop}, an overall increasing in the current.
This further complication is beyond the scope of the present work.
We conclude that results of Fig. \ref{fig:IVGS_temp} underestimate the room temperature current.

It is important to note that the barrier increasing with temperature is emphasized in present case of the simulated device by the fact that our transistor always operates in a tunneling regime, due to the absence of a ballistic contribution to the current given by holes injection over the top of the barrier.
Since no scattering states delocalized on the whole system are filled (corresponding to energies in the ballistic injection interval), the local non-equilibrium charge along the channel results to be extremely sensitive with respect to temperature variations.

The temperature dependence of the subthreshold swing $S$ has been used in the literature to distinguish between tunneling and thermal emission contributions to current in both SB- and bulk-modulated nanotube transistors \cite{Appenzeller_Therm,Appenzeller_BTB}.
In particular, a strong dependence of $S$ on temperature has been found for the thermal emission mechanism, while an almost temperature insensitive subthreshold slope results when only the tunneling current contribution is involved.
The device we have simulated shows a subthreshold swing which changes from $75$ mV/dec at $0$ K to $78$ mV/dec at $300$ K. 
This small variation completely agrees with the fact that, in our case, the current is dominated by tunneling.

\section{Conclusion}

In this work we have reported DFT, atomistic simulations of the non-equilibrium properties of bulk-modulated CNTFETs.
By tacking into account the screening properties of the CNT in term of a DFT-LDA exchange correlation functional, we have found that an unconventional over-screening situation can be generated, which has to be taken into account by means of a negative value for the nanotube quantum capacitance.
Although we have reported here the special case of the CNT (7,0), our results concerning one-dimensional screening can be applied to all small diameter tubes, where the exchange energy contribution is expected to be significant. 

The operational regimes of a short-channel nanotube transistor have been described in detail, pointing out the main differences with respect to standard silicon nano-MOSFETs.
Current saturation mechanisms in the output characteristics have been found to be critically affected by the short length and the quasi-one-dimensional screening properties of the nanotube channel.
The drain and gate capability of controlling the potential barrier have been defined in term of a simple capacit
The effects of both the system geometry and the negative quantum capacitance in addressing a well tempered design for the CNTFET have been discussed, emphasizing the worsening of the device performances which follows if the over-screening regime is entered.  
The exceptionally performing behavior expected for bulk-modulated CNTFETs has been confirmed by simulating device trans-characteristics. 
These results are quite general and hold for a generic bulk-modulated device, regardless of the over-screening problem. 

An analysis of the temperature effects on the transistor behavior has been finally addressed, pointing out that the very good transport properties of the device are preserved even at room temperature.
Our results elucidate the fundamental mechanisms which govern bulk-modulated current switching in nanotube transistors, confirming the predominant role which such devices can play in future nanoelectronics.



\section*{Acknowledgment}

We acknowledge the italian MIUR (Ministero dell'Istruzione, dell'Universit{\`a} e della Ricerca) for the financial support provided to this work within the project PRIN2004.


\begin{thebibliography}{10}
\providecommand{\url}[1]{#1}
\csname url@rmstyle\endcsname
\providecommand{\newblock}{\relax}
\providecommand{\bibinfo}[2]{#2}
\providecommand\BIBentrySTDinterwordspacing{\spaceskip=0pt\relax}
\providecommand\BIBentryALTinterwordstretchfactor{4}
\providecommand\BIBentryALTinterwordspacing{\spaceskip=\fontdimen2\font plus
\BIBentryALTinterwordstretchfactor\fontdimen3\font minus
  \fontdimen4\font\relax}
\providecommand\BIBforeignlanguage[2]{{%
\expandafter\ifx\csname l@#1\endcsname\relax
\typeout{** WARNING: IEEEtran.bst: No hyphenation pattern has been}%
\typeout{** loaded for the language `#1'. Using the pattern for}%
\typeout{** the default language instead.}%
\else
\language=\csname l@#1\endcsname
\fi
#2}}

\bibitem{Chen}
J.~Chen, C.~Klinke, A.~Afzali, and P.~Avouris, \emph{Appl. Phys. Lett.},
  vol.~86, p. 123108, 2005.

\bibitem{Lin_condMat}
Y.~M. Lin, J.~Appenzeller, J.~Knoch, and P.~Avouris, \emph{cond-mat/0501690},
  2005.

\bibitem{McEuen}
P.~L. McEuen, M.~S. Fuhrer, and H.~Park, \emph{IEEE Trans. Nanotechnol.},
  vol.~1, p.~78, 2002.

\bibitem{Appenzeller04}
J.~Appenzeller, J.~Knoch, V.~Derycke, R.~Martel, S.~Wind, and P.~Avouris,
  \emph{Rep.\ Prog.\ Phys.}, vol.~67, p.~1, 2004.

\bibitem{Yao}
Z.~Yao, C.~L. Kane, and C.~Dekker, \emph{Phys. Rev. Lett.}, vol.~84, p. 2941,
  2000.

\bibitem{Park}
J.~Park, S.~Rosenblatt, Y.~Yaish, V.~Sazanova, H.~Ustunel, S.~Braig, T.~A.
  Arias, P.~W. Brouwer, and P.~L. McEuen, \emph{NanoLett.}, vol.~4, p. 517,
  2004.

\bibitem{Tans}
S.~T. Tans, A.~R.~M. Verschueren, and C.~Dekker, \emph{Nature}, vol. 393,
  p.~49, 1998.

\bibitem{Heinze}
S.~Heinze, J.~Tersoff, R.~Martel, V.~Derycke, J.~Appenzeller, and P.~Avouris,
  \emph{Phys. Rev. Lett}, vol.~89, p. 106801, 2002.

\bibitem{Radosav}
M.~Radosavljevic, S.~Heinze, J.~Tersoff, and P.~Avouris, \emph{Appl. Phys.
  Lett.}, vol.~83, p. 2435, 2003.

\bibitem{Guo_SB}
J.~Guo, S.~Datta, and M.~Lundstrom, \emph{IEEE Trans. Electron Devices},
  vol.~51, p. 172, 2004.

\bibitem{Lin}
Y.~M. Lin, J.~Appenzeller, and P.~Avouris, \emph{Nano Lett.}, vol.~4, p. 947,
  2004.

\bibitem{Review}
A.~Pecchia and A.~Di~Carlo, \emph{Rep.\ Prog.\ Phys.}, vol.~67, p.~1, 2004.

\bibitem{Book}
A.~Di~Carlo, A.~Pecchia, L.~Latessa, T.~Fraunheim, and G.~Seifert,
  \emph{Introducing molecular electronics}.\hskip 1em plus 0.5em minus
  0.4em\relax New York: Springer-Verlag, 2005.

\bibitem{DiCarlo}
A.~Di~Carlo, M.~Gheroghe, P.~Lugli, M.~Sternberg, G.~Seifert, and
  T.~Frauenheim, \emph{Physica B}, vol. 314, p.~86, 2002.

\bibitem{Slater}
J.~C. Slater and G.~F. Koster, ``Simplified lcao method for the periodic
  potential problem,'' \emph{Phys. Rev.}, vol.~94, p. 1498, 1954.

\bibitem{Elstner}
M.~Elstner, D.~Porezag, G.~Jungnickel, J.~Elsner, M.~Haugk, T.~Frauenheim,
  S.~Suhai, and G.~Seifert, ``Self-consistent-charge density-functional
  tight-binding method for simulations of complex materials properties,''
  \emph{Phys.\ Rev.\ B}, vol.~58, p. 7260, 1998.

\bibitem{Fraunheim}
T.~Frauenheim, G.~Seifert, M.~Elstner, T.~Niehaus, C.~Koehler, M.~Amkreutz,
  M.~Sternberg, Z.~Hajnal, A.~Di~Carlo, and S.~Suhai, \emph{J. Phys.: Condens.
  Matter}, vol.~14, p. 3015, 2002.

\bibitem{Datta}
S.~Datta, \emph{Electronic Transport in Mesoscropic Systems}.\hskip 1em plus
  0.5em minus 0.4em\relax Cambridge: Cambridge University Press, 1995.

\bibitem{Brandbyge}
M.~Brandbyge, J.~L. Mozos, P.~Ordejon, J.~Taylor, and K.~Stokbro,
  ``Density-functional method for nonequilibrium electron transport,''
  \emph{Phys. Rev. B}, vol.~65, p. 165401, 2002.

\bibitem{Kadanoff}
L.~P. Kadanoff and G.~Baym, \emph{Quantum Statistical Mechanics}.\hskip 1em
  plus 0.5em minus 0.4em\relax Menlo Park: Benjamin, 1962.

\bibitem{Keldysh}
L.~V. Keldysh, \emph{Sov. Phys. JEPT}, vol.~20, p. 1018, 1965.

\bibitem{Danielewicz}
P.~Danielewicz, \emph{Ann. Phys.}, vol. 152, p. 239, 1984.

\bibitem{Tersoff}
F.~Leonard and J.~Tersoff, \emph{Phys.\ Rev.\ Lett.}, vol.~83, p. 5174, 1999.

\bibitem{Luryi}
S.~Luryi, ``Quantum capacitance devices,'' \emph{Appl.\ Phys.\ Lett.}, vol.~52,
  p. 501, 1988.

\bibitem{Buttiker}
M.~B{\"u}ttiker, ``Capacitance, admittance, and rectification properties of
  small conductors,'' \emph{J.\ Phys.:\ Condens.\ Matter}, vol.~5, p. 9361,
  1993.

\bibitem{Latessa}
L.~Latessa, A.~Pecchia, A.~Di~Carlo, and P.~Lugli, ``Negative quantum
  capacitance of gated carbon nanotubes,'' \emph{Phys. Rev. B}, vol.~72, p.
  035455, 2005.

\bibitem{Fogler}
M.~M. Fogler, \emph{Phys. Rev. Lett.}, vol.~94, p. 056405, 2005.

\bibitem{Dolgov}
O.~V. Dolgov, D.~A. Kirzhnits, and E.~G. Maksimov, \emph{Rev. Mod. Phys.},
  vol.~53, p.~81, 1981.

\bibitem{Eisenstein}
J. P. Eisenstein, L. N. Pfeiffer and K. W. West, Phys. Rev. Lett. {\bf 68}, 674
  (1992); Phys. Rev. B {\bf 50}, 1760 (1994).

\bibitem{Latessa_JCE_I}
L. Latessa, A. Pecchia and A. Di Carlo, ``One-dimensional Screening Effects in
  Bulk-modulated Carbon Nanotube Transistors'', accepted for J. Comp. Electr.

\bibitem{Fetter}
A.~L. Fetter and J.~D. Walecka, \emph{Quantum Theory of Many particle
  Systems}.\hskip 1em plus 0.5em minus 0.4em\relax New York: McGraw-Hill, 1971.

\bibitem{Calmels95}
L.~Calmels and A.~Gold, \emph{Phys.\ Rev.\ B}, vol.~51, p. 8426, 1995.

\bibitem{Venugopal}
R.~Venugopal, Z.~Ren, and M.~S. Lundstrom, \emph{IEEE Trans. Nanotech.},
  vol.~2, p. 135, 2003.

\bibitem{Natori}
K.~Natori, \emph{J. Appl. Phys.}, vol.~76, p. 4879, 1994.

\bibitem{Latessa_JCE}
L.~Latessa, A.~Pecchia, A.~Di~Carlo, and P.~Lugli, \emph{J. Comp. Electr.},
  vol.~4, p.~51, 2005.

\bibitem{Appenzeller_BTB}
J.~Appenzeller, Y.~M. Lin, J.~Knoch, and P.~Avouris, \emph{Phys. Rev. Lett.},
  vol.~93, p. 196805, 2004.

\bibitem{Appenzeller_Therm}
J.~Appenzeller, M.~Radosavljevic, J.~Knoch, and P.~Avouris, \emph{Phys. Rev.
  Lett.}, vol.~92, p. 048301, 2004.

\end{thebibliography}







\end{document}